\documentclass[aps,prl,preprint,superscriptaddress]{revtex4-1}
\def\pref#1{(\ref{#1})}
\usepackage{graphicx} 
\usepackage{amssymb}

\begin{document}

\preprint{APS/123-QED}


\title{Ultra-Strong Light-Matter Coupling Regime with Polariton Dots}

\author{Y. Todorov}\email{yanko.todorov@univ-paris-diderot.fr}
\affiliation{Laboratoire "Mat{\'e}riaux et Ph{\'e}nomenes
Quantiques", Unversit{\'e} Paris Diderot-Paris 7, CNRS-UMR 7162,
75013 Paris, France}
\author{A. M. Andrews}
\affiliation{Solid State Electronics Institute TU Wien, Floragasse
7, A-1040 Vienna, Austria}
\author{R. Colombelli}
\affiliation{Institut d'Electronique Fondamentale, Univ. Paris-Sud
and CNRS-UMR 8622, F-91405 Orsay, France}
\author{S. De Liberato}
\affiliation{Laboratoire "Mat{\'e}riaux et Ph{\'e}nomenes
Quantiques", Unversit{\'e} Paris Diderot-Paris 7, CNRS-UMR 7162,
75013 Paris, France} \affiliation{Department of Physics,
University of Tokyo, Bunkyo-ku, Tokyo 113-0033, Japan}
\author{C. Ciuti}
\affiliation{Laboratoire "Mat{\'e}riaux et Ph{\'e}nomenes
Quantiques", Unversit{\'e} Paris Diderot-Paris 7, CNRS-UMR 7162,
75013 Paris, France}
\author{P. Klang}
\affiliation{Solid State Electronics Institute TU Wien, Floragasse
7, A-1040 Vienna, Austria}
\author{G. Strasser}
\affiliation{Solid State Electronics Institute TU Wien, Floragasse
7, A-1040 Vienna, Austria}
\author{C. Sirtori}
\affiliation{Laboratoire "Mat{\'e}riaux et Ph{\'e}nomenes
Quantiques", Unversit{\'e} Paris Diderot-Paris 7, CNRS-UMR 7162,
75013 Paris, France}


\date{\today}

\begin{abstract}
The regime of ultra-strong light-matter interaction has been
investigated theoretically and experimentally, using
zero-dimensional electromagnetic resonators coupled with an
electronic transition between two confined states of a
semiconductor quantum well. We have measured a splitting between
the coupled modes that amounts to 48{\%} of the energy transition,
the highest ratio ever observed in a light-matter coupled system.
Our analysis, based on a microscopic quantum theory, shows that
the non-linear polariton splitting, a signature of this regime, is
a dynamical effect arising from the self-interaction of the
collective electronic polarization with its own emitted field.
\end{abstract}


\maketitle

Light-matter interaction in the strong coupling regime is a
reversible process in which a photon is absorbed and reemitted by
an electronic transition at a rate equal to the coupling energy
divided by the Plank constant $\hbar$. This situation is observed
in systems where an electronic transition, embedded in an optical
cavity, has the same energy as the confined photonic mode
\cite{Adreani_2003, Yamamoto_2000}. An adequate description of the
system is given using quantum mechanics that permits to describe
the stationary states as mixed particles, the polaritons, which
are a linear combination of light and matter wavefunctions
\cite{Hopfield_1958, Artoni_Birman_1991}. Recently, the cavity
polaritons produced by the intersubband transitions in a highly
doped quantum well have received considerable interest
\cite{Dini_2003, Todorov_PRL_2009, Gunter_Nature_2009}. In this
system the number of available excitations per unit volume can be
very high, and an unexplored limit can be reached where the
interaction energy $\hbar \Omega_R $ ($\Omega_R$ is the Rabi
Frequency) is of the same order of magnitude as the transition
$\hbar \omega_{12}$, the recently named "ultra strong coupling
regime" \cite{Ciuti_PRB_2005}. We have developed a microscopic
quantum theory and provided experimental evidences linking this
regime to the collective phenomena in the confined electronic gas
\cite{Ando_1982}. Our studies are conducted with "polaritons dots"
produced using zero-dimensional microcavities. In this systems we
have measured an unprecedent ratio $2 \Omega_R/\omega_{12} =
0.48$, which is more than twice the highest values reported in the
literature \cite{Todorov_PRL_2009, Gunter_Nature_2009,
T_Niemczyk_2010}. This high ratio has allowed us to observe, for
the first time, the nonlinearities in the coupling constant, which
are indisputable features of the ultra-strong coupling regime.

\begin{figure}
\includegraphics[scale=0.42]{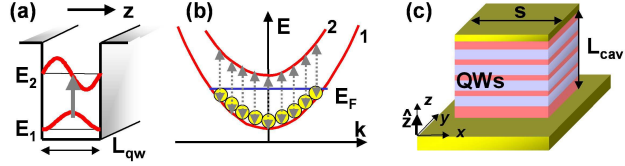}
\caption {a) Semiconductor quantum well (QW) with two subbands of
energies $E_1$ and $E_2$.  b) In-plane parabolic dispersion of the
energy subbands, with a sketch of the action of the polarization
operator $b^\dagger$ (arrows). $E_F$ is the Fermi level. c)
Multiple QWs embedded in a square shaped microcavity.}
\label{Fig1}
\end{figure}

A natural framework to describe the interaction of quantized light
with a solid state system is the multipolar coupling Hamiltonian
involving the electric displacement $\mathbf{D}(\mathbf{r})$
\cite{CCT_b2001, Welsch_b2001}, and the local polarization field
$\mathbf{P}(\mathbf{r})$ of the optically active excitations
\cite{Haug_Koch_b2004}. In the case of an intersubband transition
of energy $\hbar \omega_{12}$ we can define a polarisation
operator as $b_\mathbf{q}^\dagger=1/\sqrt{N}\sum_\mathbf{k}
c^\dagger_{2\mathbf{k}+\mathbf{q}}c_{1\mathbf{k}}$ where
$\mathbf{q}$ is the photon wavevector, $N=N_1-N_2$ is the
population difference between the subbands 1 and 2, and
$c_{1\mathbf{k}}$ and $c^\dagger_{2\mathbf{k}}$ are the
corresponding fermionic destruction/creation operators
\cite{Ciuti_PRB_2005}. Fig. \ref{Fig1}(a) illustrates a quantum
well (QW) of thickness $L_{qw}$ and Fig. \ref{Fig1}(b) the
schematic principle of the operator $b_\mathbf{q}^\dagger$ which
promotes coherently each electron from the ground subband to the
excited one, with an identical probability of $1/\sqrt{N}$. The
polarisation density is then
$\mathbf{P}(z,\mathbf{r}_{\parallel})=(\mathbf{\hat{z}}d_{12}\sqrt{N}/L_{qw}S)
\sum_\mathbf{q}(b_\mathbf{q}^\dagger+b_\mathbf{q})
\sqrt{2}\cos(\mathbf{q.r}_{\parallel})\eta (z)$, where $z$ is the
well growth axis, $\mathbf{\hat{z}}$ the corresponding unit vector
and $\mathbf{r}_{\parallel}=(x,y)$ the in-plane position (Fig.
\ref{Fig1}(c)), $d_{12}$ the transition dipole and $S$ the area of
the system \cite{ODiStephano_1999}. The function $\eta (z)$ equals
1 in the quantum well slab and 0 everywhere else. Let the
polarization field interact with the fundamental TM mode of 0D
square-shaped microcavity of frequency $\omega_c$ and dimensions
$s \times s \times L_{cav}$ (Fig. \ref{Fig1}(c)). The electric
displacement of this mode is
$\mathbf{D}(\mathbf{r})=\mathbf{\hat{z}}i\sqrt{\varepsilon
\varepsilon_0 \hbar \omega_c/L_{cav} S}(a^\dagger-a)\cos(\pi
y/s)$, with $\varepsilon$ the background dielectric constant, and
$a^\dagger$ the photonic creation operator \cite{Todorov_Opex2010,
Kakazu_Kim_1994}. We have therefore $|\mathbf{q}|=\pi/s$ which is
very small compared to the typical electron wavevectors
$\mathbf{k}$ and will be neglected in the definition of
$b_\mathbf{q}^\dagger$. In this long wavelength limit the dipole
interaction Hamiltonian $H_I = \int \mathrm{d}^3 r (-\mathbf{D
\centerdot P} + \mathbf{P}^2/2)/\varepsilon \varepsilon_0$
\cite{CCT_b2001, Welsch_b2001} can be expressed as:
\begin{equation}\label{H_I}
H_I = id_{12}\sqrt{\frac{N\hbar \omega_c} {2\varepsilon
\varepsilon_0 L_{cav} S}}(a-a^\dagger)(b+b^\dagger)+\frac{d_{12}^2
N}{\varepsilon \varepsilon_0 L_{qw} S}(b+b^\dagger)^2
\end{equation}

The quadratic term of \pref{H_I} describing the polarization
self-interaction is usually disregarded, but we are going to show
that it plays an important role in the limit of very strong
light-matter coupling. The notations of the problem can be greatly
simplified introducing the \textit{plasma frequency} $\omega_P$:
\begin{equation}\label{wP}
\omega_P^2 = \frac{e^2f_{12} N}{m^\ast\varepsilon \varepsilon_0
L_{qw} S} =\frac{2 \omega_{12} d^2_{12} N}{\hbar \varepsilon
\varepsilon_0 L_{qw} S}
\end{equation}

where $f_{12} = 2m^\ast\omega_{12}d^2_{12}/\hbar e^2$ is the
oscillator strength, and $m^\ast$ the effective electron mass
\cite{Ando_1982}. The full Hamiltonian of the system, including
the contributions of the cavity and the material excitation is
then:
\begin{eqnarray}\label{H1}
H = \hbar \omega_c(a^\dagger a + 1/2) +\hbar \omega_{12} b^\dagger b \nonumber\\
+\frac{i\hbar
\omega_P}{2}\sqrt{f_w\frac{\omega_c}{\omega_{12}}}(a-a^\dagger)(b+b^\dagger)
+\frac{\hbar \omega_P^2}{4\omega_{12}}(b+b^\dagger)^2
\end{eqnarray}

Here $f_w=L_{qw}/L_{cav}$ is the overlap factor between the
polarization medium and the cavity mode. Clearly, the polarization
self-energy depends only on the matter frequencies $\omega_{12}$,
$\omega_P$. The matter part $H_{pol}= \hbar \omega_{12} b^\dagger
b + \hbar \omega_P^2/4\omega_{12}(b+b^\dagger)^2$ can therefore be
diagonalized separately through the Bogoliubov procedure
\cite{Schwabl_b1997}, by defining a destruction operator $p$ such
that $[p,H_{pol}]=\hbar \widetilde{\omega}_{12} p$. This leads to
$H_{pol} = \hbar \widetilde{\omega}_{12} p^\dagger p$ where
\begin{equation}\label{WP2}
\widetilde{\omega}_{12} =
\sqrt{\omega_{12}^2+\omega_P^2}\phantom{q}\mathrm{and}\phantom{q}
p =
\frac{\widetilde{\omega}_{12}+\omega_{12}}{2\sqrt{\widetilde{\omega}_{12}\omega_{12}}}b
+\frac{\widetilde{\omega}_{12}-\omega_{12}}{2\sqrt{\widetilde{\omega}_{12}\omega_{12}}}b^\dagger
\end{equation}

The new polarization eigenfrequency  $\widetilde{\omega}_{12} =
\sqrt{\omega_{12}^2+\omega_P^2}$ is identical to the result of
Ando et al. \cite{Ando_1982}, describing the collective
oscillations of two dimensional electrons, an effect known as the
"depolarization shift" \cite{Helm_b2000}. This phenomenon appears
naturally from the complete interaction Hamiltonian \pref{H1}
expressed in the Power-Zienau-Woolley (PZW) representation
\cite{CCT_b2001}.

Moreover, it is remarkable that the coupling with the cavity mode
is also proportional to $\omega_P$. Therefore the limit of very
strong light-matter interaction also implies a large
depolarization shift. Our model allows us to clearly quantify the
link between the two features. Using \pref{WP2} we replace
$b^\dagger/b$ by the renormalized polarization operators
$p^\dagger/p$, to obtain a linear interaction Hamiltonian:
\begin{eqnarray}\label{Hplasmon}
H =\hbar \omega_c(a^\dagger a + 1/2) +\hbar
\widetilde{\omega}_{12} p^\dagger p \nonumber\\ +i\frac{\hbar
\omega_P}{2}\sqrt{f_w\frac{\omega_c}{\widetilde{\omega}_{12}}}(a-a^\dagger)(p+p^\dagger)
\end{eqnarray}

The eigenvalues of $\pref{Hplasmon}$ are provided by the equation:
\begin{equation}\label{Eigen}
(\omega^2-\widetilde{\omega}_{12}^2)(\omega^2 -\omega_c^2) =
f_w\omega_P^2\omega_c^2
\end{equation}

The equations \pref{Hplasmon} and \pref{Eigen} describe the
coupling between two independent oscillators: the bare microcavity
mode and the bosonic excitation renormalized by its own radiated
field. Its roots, $\omega_{UP}$ and $\omega_{LP}$ are the
frequencies of the two polariton states. Note that the relevant
features of the ultra-strong coupling are present trough the
anti-resonant terms $a^\dagger p^\dagger$ and $ap$. The
Hamiltonian \pref{Hplasmon} can be related to the full standard
minimal coupling Hamiltonian used so far for the study of the
ultra-strong coupling regime \cite{Ciuti_PRB_2005}, through the
inverse PZW transformation \cite{Add_Mat}.

Eq. \pref{Eigen}, which is the polariton dispersion relation
allows to introduce an effective dielectric constant through the
usual relation
$\bar{\varepsilon}(\omega)\omega^2/c^2=\mathbf{q}^2=\varepsilon
\omega_c^2/c^2$. Identifying $\bar{\varepsilon}(\omega)$ from
\pref{Eigen} we find:
\begin{equation}
1/\bar{\varepsilon}(\omega) =
f_w/\varepsilon_{qw}(\omega)+(1-f_w)/\varepsilon
\end{equation}

Here
$\varepsilon_{qw}(\omega)=\varepsilon(1-\omega_P^2/(\omega^2-\omega_{12}^2))$
is the usual QW slab dielectric constant \cite{Wendler_1996}. This
is precisely the results of Zaluzny et al. \cite{Zaluzny_1999},
which confirms the pertinence of our model. Moreover, in the limit
$f_w=1$ we recover the homogeneous Hopfield model
\cite{Hopfield_1958, Artoni_Birman_1991}.

Taking the resonant case, $\omega_c=\omega_{12}$, from
\pref{Eigen} we can deduce that, \textit{up to third order} in
$\omega_P/\omega_{12}$, the polariton splitting is
$\omega_{UP}-\omega_{LP} = \sqrt{f_w}\omega_P = 2\Omega_R $. Note
that the Rabi frequency $\Omega_R$ goes to zero when $f_w$ does
it, independently from the value of $\omega_P$. Indeed, by
changing the overlap factor in \pref{Hplasmon} and \pref{Eigen}
one can move the system from the ultra-strong coupling regime to
the uncoupled situation. Both situations, however, bring the
signatures of the plasma frequency $\omega_P$, which appears as
the fundamental quantity for the light-coupled 2D electronic
system.

\begin{figure}
\includegraphics[scale=0.42]{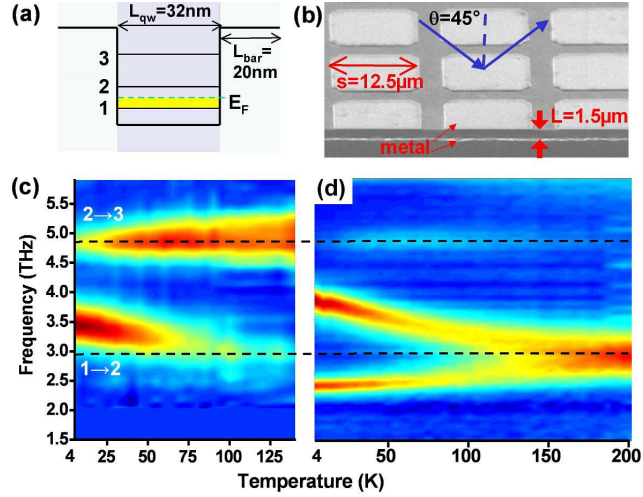}
\caption {a) Schematics of the QW media of our samples. The first
two electronic transitions are $E_2-E_1 = 12.4$ meV (3 THz) and
$E_3-E_2 = 19.9$ meV (4.8 THz). b) Electronic microscope image of
the cleaved facet of an array of metal-dielectric-metal patch
cavities. c) Contour plot of the QW multipass absorption as a
function of the frequency and temperature, revealing the first
$1\rightarrow 2$ and the second $2\rightarrow 3$ QW transitions.
d) Reflectivity contour plot, with $\theta=45^\circ$, as a
function of frequency and temperature, for a cavity ($s=12.5$
$\mu$m) with $\omega_c = 3$ THz, resonant with the $1\rightarrow
2$ QW transition.} \label{Fig2}
\end{figure}

To confirm experimentally these effects, it is necessary to have a
system with a large ratio $\omega_P/\omega_{12}$. This situation
is readily obtained for intersubband transition in the THz
spectral region using highly doped quantum wells. Our system is
composed of a thin
$\mathrm{GaAs}/\mathrm{Al}_{0.15}\mathrm{Ga}_{0.85}\mathrm{As}$
multi-quantum well structure represented in Fig. \ref{Fig2}(a). It
comprises $N_{qw} = 25$ quantum wells of width $L_{qw} = 32$nm
separated by $L_{bar} = 20$nm barriers, silicon $\delta$-doped
with a sheet density of $2\times 10^{11}$cm$^{-2}$. The
intersubband transition energies between the first three subbands
are respectively $E_{12} = E_2 - E_1= 12.4$ meV ($3$ THz) and
$E_{32} = E_3 - E_2 =19.9$ meV ($4.8$ THz). Light is confined in
the vertical direction by a metal plate on one side and a
two-dimensional array of squared metallic pads of size $s$ on the
other. In the lateral direction the confinement is provided by the
strong impedance mismatch between the regions covered and
uncovered by the metal, as shown in Fig. \ref{Fig2}(b)
\cite{Todorov_PRL_2009, Todorov_Opex2010}. This arrangement
creates a 0D cavity with a reduced volume with respect to the
wavelength of the mode, on the order of $V_{cav}/\lambda_{res}^3=
10^{-4}$. The optical response of the cavities is studied in
spectrally resolved reflectometry measurements, at different
angles of incidence, using a Bruker Fourier Transform Infrared
Spectrometer \cite{Todorov_Opex2010}. In Fig. \ref{Fig2}(c,d) we
present two contour plots as a function of the temperature: the
absorption of a reference uncoupled system (\ref{Fig2}(c)) and the
reflectivity of a resonantly coupled cavity, $\hbar \omega_c =
E_{12}$ (\ref{Fig2}(d)). In the reference, the absorption is
measured in a multipass configuration using a sample with no
cavity. At high temperature electrons are populating several
subbands and have a similar density on the fundamental state,
$E_1$, and on the first excited state $E_2$. When the temperature
is lowered, the population of $E_1$ increases with respect to
$E_2$, which allows to vary $N=N_1-N_2$ and $\omega_P$. In Fig.
\ref{Fig2}(c) one can clearly observe the two absorption peaks
corresponding to $E_{12}$ and $E_{23}$. Their intensities have
opposite behaviour as a function of the temperature, as expected
by the electron redistribution on the subbands. Moreover, the
$E_{12}$ transition has blue shifted at low temperatures due to
the increased electronic density on the ground state, which is the
evidence of the depolarization shift. In Fig. \ref{Fig2}(d) the
reflectivity peak clearly splits into the upper and lower
polariton branches for temperature below 120 K. The separation
between the branches keeps increasing down to 4 K, up to a maximum
value of 1.41 THz. In the same panel, we can observe also the
$E_{23}$ transition that peaks between 60 K and 80 K and
disappears at 4 K.

\begin{figure}
\includegraphics[scale=0.42]{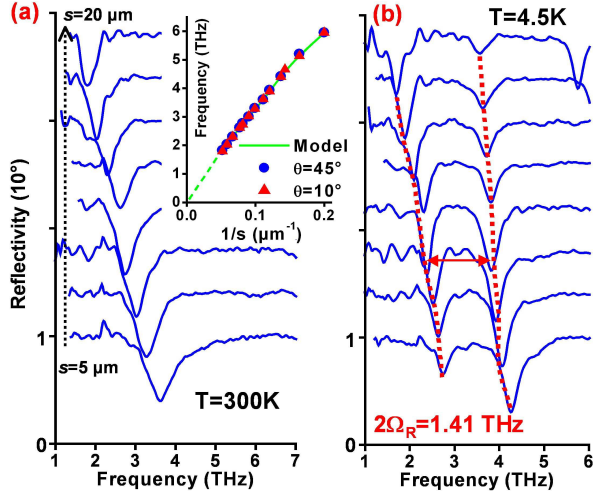}
\caption {a) Reflectivity spectra for different cavities with
decreasing size $s$, for an incident angle $\theta=10^\circ$.
Spectra have been offset for clarity. The inset summarizes the
frequency of the cavity mode as a function of $1/s$ measured at
$10^\circ$ (red triangles) and $45^\circ$ (blue dots) angles of
incidence. b) Low temperature ($T$=4.5 K) reflectivity spectra
$\theta=10^\circ$, for the same cavities as in a). The minimal
polariton separation is the Rabi splitting $2\Omega_R = 1.41$
THz.} \label{Fig3}
\end{figure}

Figure \ref{Fig3} presents a detailed characterization of the
cavities with different patch widths $s$. At $T$=300 K the
reflectivity spectra presented in Fig. \ref{Fig3}(a) feature only
the bare cavity mode allowing us to map its frequency $\omega_c$
as a function of $1/s$ (inset). As shown in the inset, $\omega_c$
is independent from the incident angle $\theta$, attesting the 0D
character of the microcavities \cite{Todorov_Opex2010}. Fig.
\ref{Fig3}(b) summarizes the spectra at $T$=4.5 K, where the
cavity mode is coupled with the fundamental intersubband
transition $1\rightarrow 2$, with a very clear anticrossing
behaviour. The remarkable feature of our data is the minimum
polariton separation, the Rabi splitting, $2\Omega_R=
\omega_{UP}-\omega_{LP}= 1.41$ THz, which is 48\% of the bare
intersubband transition $E_{12}=3$ THz. This value is, to our
knowledge, the largest fraction ever measured in a light-matter
interacting system. Notice that the typical width of the
resonances is in the order of 250 GHz, thus a factor of 5 to 6
less than the Rabi splitting.

\begin{figure}
\includegraphics[scale=0.42]{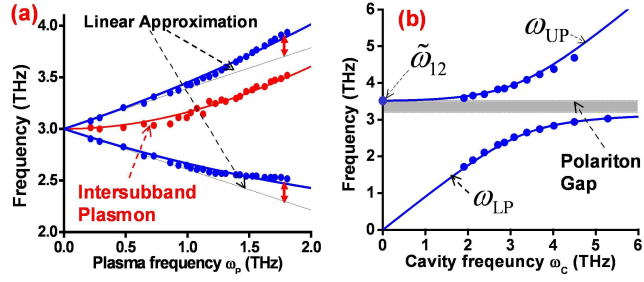}
\caption {a) The two polariton peaks $\omega_{UP}$  and
$\omega_{LP}$ (in blue) as a function of the plasma frequency
$\omega_P$, in the resonant case $\omega_c=\omega_{12}$. In red we
have plotted the energy of the intersubband plasmon, as derived
from the absorption experiment (dots) and Eq. \pref{wP}
(continuous line). b) Polaritons frequencies  $\omega_{UP}$  and
$\omega_{LP}$ as a function of the cavity frequency $\omega_c$.
The dots are experimental results, and the continuous lines are
the roots of Eq. \pref{Eigen} with $f_w = 0.62$ and $\omega_P
=1.8$THz. The hatched zone indicates the polariton gap.}
\label{Fig4}
\end{figure}

In Fig. \ref{Fig4}(a) the polariton peak energies (blue dots) for
the resonant case (Fig. \ref{Fig2}(c)) are plotted as a function
of $\omega_P$, which can be easily inferred from the polariton
splitting ($2\Omega_R=\sqrt{f_w}\omega_P$) and the knowledge of
$f_w$: for our structure we have $f_w =
N_{qw}L_{qw}/L_{cav}=0.62$. On the same graph, we can therefore
plot the $1\rightarrow 2$ peaks of the uncoupled structure as a
function of the temperature, measured from the multipass
absorption spectra. The solid blue lines are the roots of equation
\pref{Eigen} at resonance when $\hbar \omega_c = E_{12}$, while
the red line corresponds to the case $f_w = 0$. The agreement with
the data, using no adjustable parameters, is excellent. Moreover,
the strong deviation of the polariton curves from the linear
approximation $\omega = \omega_{12}\pm \sqrt{f_w}\omega_P/2$,(red
arrows in Fig. \ref{Fig4}(a)), is unambiguous evidence of the
ultra-strong coupling regime \cite{Ciuti_PRB_2005}.

A relevant consequence of the the large ratio
$2\Omega_R/\omega_{12}$ is the opening of an energy gap, $\Delta
E_{gap}$, where no polaritonic solutions can be found. This is
illustrated in the Fig. \ref{Fig4}(b) where the polariton
resonances are plotted as a function of the cavity frequency
$\omega_c$, both from the experiment and the roots of Eq.
\pref{Eigen}. The forbidden frequencies correspond to destructive
interference between the electromagnetic field radiated by the
electronic oscillations and the bare microcavity photon field.
This is analogue to the case of the forbidden optical phonon band
of bulk polar semiconductors \cite{Kittel_b1963}. The evidence of
the forbidden band, $\Delta E_{gap} \approx
f_w\omega_P^2/2\omega_{12}= 2\Omega_R^2/\omega_{12}$ is another
proof of the strength of the light-matter coupling. From our
measurements we deduce $\Delta E_{gap}=330$ GHz, which is greater
than the polariton linewidth (250 GHz). Such a gap has already
been observed for bulk ($f_w=1$) excitonic systems, but never for
any microcavity-coupled electronic system, to our knowledge.

In conclusion, we have explored a 0D microcavity coupled high
density electronic system which has allowed us to reach the
ultra-strong light-matter coupling regime. Our results, both
theoretical and experimental, show that in this limit light-matter
interaction is linked to the collective excitations of the
electron gas, which yield the dominant non-linearity in the
polariton splitting. This occurs because for high electronic
densities the energy exchanged by the electronic polarisation with
its own emitted field is a non-negligible effect. Theoretically,
this is expressed by the weight of the quadratic term that, in the
dipolar Hamiltonian, becomes comparable to the light-matter
interaction one. We believe these results form the basis for
quantum devices based on the ultra-strong light-matter coupling in
the THz/$\mu$-wave spectral range.

We thank L. Tosetto and H. Detz for technical help. We gratefully
acknowledge support from the French National Research Agency
through the program ANR-05-NANO-049 Interpol, the ERC grant "ADEQUATE" and from the Austrian
Science Fund (FWF).

\end{document}